# Pulsed magnetic field generation system for laser-plasma research

A.G.Luchinin[1], V.A.Malyshev[1], E.A.Kopelovich[1], K.F.Burdonov[1], M.E.Guschin[1], M.V.Morozkin[1], M.D.Proyavin[1], R.M.Rozental[1], A.A.Soloviev[1], M.V.Starodubtsev[1], A.P.Fokin[1], J.Fuchs[1,2], and M.Yu.Glyavin[1]

[1]*Institute of Applied Physics RAS (IAP RAS), Nizhny Novgorod, Russia*
[2]*Sorbonne Université, Ecole Polytechnique, Institut Polytechnique de Paris (LULI - CNRS), Paris, France*
glyavin@appl.sci-nnov.ru

**Abstract**

An up to 15 T pulsed magnetic field generator in a volume of a few cubic centimeters has been created for experiments with magnetized laser plasma. The magnetic field is created by a pair of coils placed in a sealed reservoir with liquid nitrogen, which is installed in a vacuum chamber with a laser target. The bearing body provides the mechanical strength of the system both in the case of co-directional and oppositely connected coils. The configuration of the housing allows laser radiation to be introduced into the working area between the coils in a wide range of directions and focusing angles, to place targets away from the symmetry axis of the magnetic system, and to irradiate several targets simultaneously.

## 1. Introduction

Pulsed sources of a strong magnetic field are widely used in various devices and systems, including laser-plasma research. Interest in these works is largely stimulated by the search for new schemes for laser-plasma generation of charged particle beams and problems of controlling their characteristics, including the collimation of charged particle beams [1, 2]. Work in the field of high energy density physics is another important application of strong magnetic fields in laser-plasma research. Here, the prospect of creating a magnetized laser plasma opens up wide experimental opportunities for studying the entire spectrum of problems, ranging from laboratory modeling of astrophysical phenomena (including the study of the formation mechanisms of astrophysical jets [3] and accretion flows [4] and the conditions for the development of MHD [5] and other types of instabilities [6] that arise during the interaction of plasma flows with a magnetic field) to the development of new directions in controlled thermonuclear fusion related to using external magnetic fields to magnetize ions and increase their temperature [7], reduce losses due to limiting the electron thermal conductivity [8], and suppress hydrodynamic instabilities [9, 10], as well as to implement other methods of increasing the efficiency of heating a thermonuclear target [11].

In this paper, we describe a pulsed magnetic system created at the IAP RAS for research in high energy density physics and laboratory astrophysics at the PEARL laser-plasma facility [12]. Section 2 defines the requirements for the magnetic system. Section 3 contains a description of its design and the results of measurements of the parameters of the magnetic system. Section 4 formulates the results of the work and provides brief data on experiments using the created magnetic system.

## 2. Justification of the parameters of the magnetic system for a high-energy laser-plasma experiment

The problems of creating an experimental platform for studying a magnetized laser plasma are related with the need to generate strong magnetic fields of a given configuration on centimeter scales. To understand the characteristic values of the magnetic field and its spatial scales, one can use the estimates made in [13, 14]. Assume that the laser plasma is instantly created at some point in space with a uniform external magnetic field $B_0$ and starts to expand isotropically at a supersonic speed $V_0$. We suppose that the total energy stored in the expanding plasma flow is equal to $E_0$. Due to the high electrical conductivity of the laser-produced plasma, the magnetic field at the initial stage of its expansion can be considered frozen into the plasma. As a result, during the expansion of the plasma cloud, the magnetic field lines move apart and the magnetic field inside the plasma cloud decreases. By the time the plasma cloud reaches its largest scale, the magnetic field inside the plasma decreases many times compared to the background value. Thus, in the cloud, one can distinguish the inner part, where the plasma is effectively diamagnetic, and the outer part, which directly interacts with the magnetic field of the experimental setup. Neglecting the effects of magnetic field penetration into the plasma, the characteristic size to which the plasma cloud can expand into an external magnetic field will be determined by the equality of the initial cloud energy and the total energy of the magnetic field displaced by the plasma. If we assume that the plasma cloud has the shape of a sphere of radius $R_b$, then the energy of the displaced magnetic field can be written as

$$\frac{4\pi}{3} R_b^3 \frac{B_0^2}{8\pi} = E_0, \qquad (1)$$

from where we obtain an estimate for the so-called classical plasma deceleration radius:

$$R_b = (6E_0/B_0^2)^{1/3} \qquad (2)$$

Figure 1 shows the relationship between the plasma deceleration radius (1) and the values of the external magnetic field induced at energies $E_0$ = 1-100 J, which are typical for laboratory laser-plasma experiments.

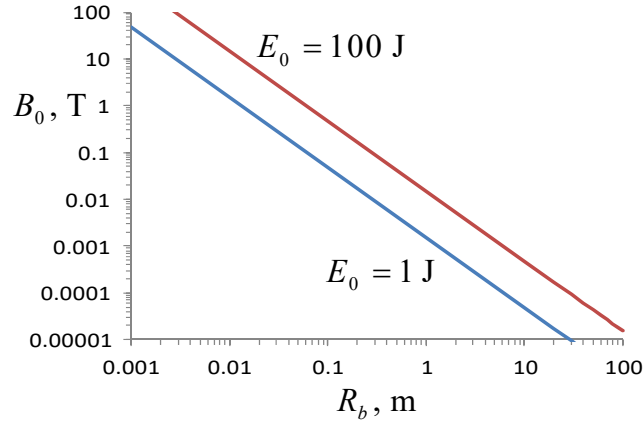

Fig.1. The classical plasma-cloud deceleration radius $R_b$ as a function of the external magnetic field $B_0$ for typical energies of the laser-plasma cloud.

A significant number of experimental studies of the interaction of plasma clouds with an external magnetic field are performed at $B_0 \sim$ 100 - 1000 G [15-17]. In this case, the characteristic dimensions of the plasma cloud lie in the range $R_b \sim$ 10 cm - 1 m and the characteristic plasma densities turn out to be of the order of $10^{14}$ cm$^{-3}$. The main methods for studying plasma dynamics under these conditions are high-speed photography and probe techniques. A feature of laser-plasma experiments in this mode is the need to place a target in a vacuum chamber with a volume of at least a few cubic meters.

In laser-plasma experiments with a high energy density in the beam, another mode, characterized by significantly higher values of the electron density, namely, $10^{17}$ - $10^{19}$ cm$^{-3}$, and several orders of magnitude smaller plasma dimensions, appears to be very convenient. This mode makes it possible to use optical methods of plasma diagnostics, primarily laser interferometry. Indeed, the characteristic phase increment

$$\Delta\varphi = \frac{\omega}{c}\Delta n \times 2R_b \sim 2\pi\frac{N}{N_{cr}}\frac{R_b}{\lambda} \tag{3}$$

in the interferometric circuit is 0.1-10 rad, which is quite convenient for measurements (here, $N_{cr} = \omega_L m/(4\pi e^2)$ is the critical plasma density for optical radiation at the frequency $\omega_L$ and wavelength $\lambda = 2\pi c/\omega_L$). At the same time, this mode requires the use of much stronger magnetic fields, of the order of several tens of tesla [18-20]. In a dense plasma, magnetic fields of 10 T or more are required to retain the key dimensionless parameters for modeling of space phenomena [4]. In particular, for a laser plasma with a density of $10^{18}$ cm$^{-3}$ and a temperature of about 30 eV the condition of strong magnetization of electrons, which is determined by the Hall parameter ($\omega_C/\nu > 10$, where $\omega_C$ – is the electron gyrofrequency and $\nu$ is the frequency of Coulomb collisions) is realized at a field level of 10 T or more. The same field values are required for setting up experiments in which the ion cyclotron radius is small compared to the scale of the cloud expansion. According to Fig. 1, for the same values of the total plasma energy, the characteristic maximum size $R_b$ of the plasma cloud is of the order of a few millimeters, i.e., experiments are possible in compact target chambers.

Obviously, in experiments on the interaction of a laser plasma with an external magnetic field, the characteristic scale $L$ of the latter should be several times larger than $R_b$ to make it possible to study not only the formation of a diamagnetic cavity, but also the dynamics of further interaction of the plasma with the magnetic field and background medium. According to Fig. 1, for the fields $B_0 \sim 10$ T the scale $L$ occupied by the magnetic field should be at least 1 cm. The characteristic time required for the plasma to reach the size $R_b$ in this case is $\tau \sim R_b/V_0 \sim 10$ ns. This simple reasoning leads to the following requirements for the parameters of the magnetic system: the magnetic field is no less than 10 T, the dimensions in all directions are no less than 1 cm, and the pulse duration of the magnetic field is much more than 10 ns. In addition, the magnetic system should provide access of high-power and diagnostic laser radiation to the target in the widest possible range of angles and directions.

An example of an experiment scheme for studying the interaction of a laser-produced plasma with an external magnetic field is shown in Fig. 2. A laser pulse (radiation wavelength ~ 1 μm, pulse duration ~ 1 ns, and energy ~ 1-100 J) irradiates the surface of a solid target. The target is located inside the coils that create a locally uniform magnetic field. In this geometry, the plasma flow flies out mainly in the direction perpendicular to the target surface. The characteristic velocity of the plasma flow in such experiments is 100 -1000 km/s.

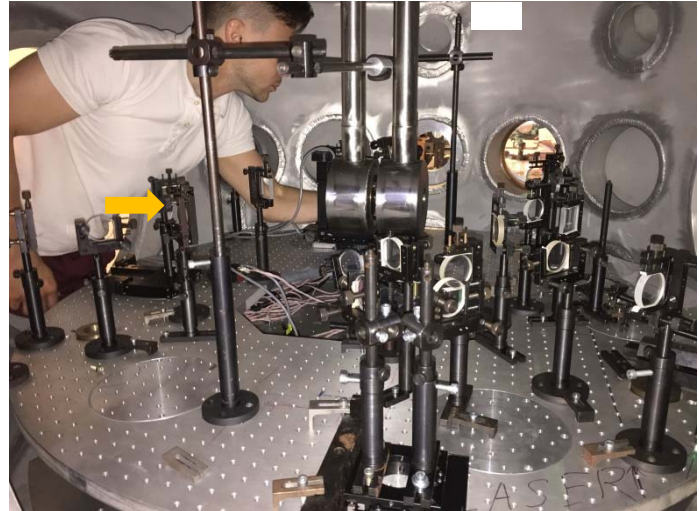
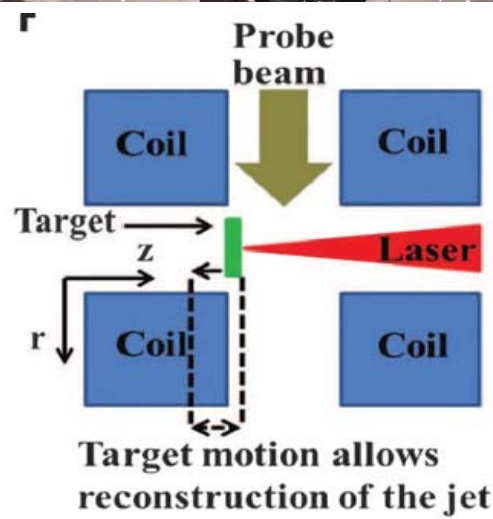

Fig. 2. A photograph of the experimental chamber (left) and the experiment scheme (right) for studying the interaction of the laser plasma flow with an external magnetic field. Horizontal arrows indicate the magnetic field direction.

The current laser-plasma experiments with magnetic fields of 10 T or more use miniature disposable coils, which, as a rule, are destroyed by the flow of a pulsed current and as a result of exposure to high-power laser radiation [21, 22]. The need to replace the coils after each shot of the laser setup creates inconveniences during the experiment. The variant of stationary coils protected from laser radiation by a metal shell and effectively cooled between shots of the laser setup seems more promising. Since laser-plasma experiments are performed in high vacuum chambers, and a typical pulse repetition rate of laser radiation is one pulse per several tens of minutes, it is expedient to use a cryostat filled with liquid nitrogen to cool the coils. As a result, the simplest stationary magnetic system can consist of a pair of coils separated by more than 1 cm and fixed in a sealed housing with double walls, between which liquid nitrogen is poured. In this case, along with the mechanical strength and vacuum tightness, the housing design should provide the following capabilities:
- setting up a target attached to the positioning system;

- input and output of a diagnostic laser beam with a diameter of at least 1 cm in two mutually perpendicular directions;
- the ability to enter high-power laser radiation in a wide range of angles with respect to the direction of the magnetic field.

The last point imposes one more restriction on the design of the magnetic system. Indeed, to create a laser-produced plasma by substance ablation from the surface of a solid target, the intensity of nanosecond laser radiation on the target surface should be of the order of $10^{11}$-$10^{15}$ W/cm$^2$. Typical parameters of a nanosecond laser beam that will be used in experiments are as follows: duration 1 ns, energy less than 100 J, and diameter 10 cm. Simple estimates show that when a beam with an energy of 10 J is used, an intensity of $10^{15}$ W/cm$^2$ is provided when the laser radiation is focused into a spot with a diameter of 30 μm, which corresponds to the focusing F/20 or better. An F/10 focusing was employed in the experiments. Thus, the structural elements of the magnetic system should not interfere with such focusing of laser radiation into the central region of the magnetic system.

The requirements for the magnetic field uniformity in the region where the laser plasma interacts with the magnetic field are not very high, since, in any case, in the course of experiments, the magnetic field is strongly disturbed by the invading plasma flow, and the process of magnetic field penetration into the plasma is accompanied by the development of instabilities at the boundary between plasma and magnetic field. When designing the magnetic system described below, we restricted ourselves to the requirements for the magnetic field uniformity of the order of $\Delta B_0 / B_0 \leq 10\%$ in the gap between coils.

The studies of the processes of interaction of a laser-produced plasma of high energy density with a magnetic field are not limited to the case of a locally uniform distribution of the magnetic field. A number of problems on laboratory modeling of space plasma phenomena require the creation of highly non-uniform magnetic fields near the target, including the points with a zero magnetic field. Such problems include modeling of plasma processes during magnetic reconnection [23] and various aspects of the dynamics of accretion disks [24]. In particular, setting up model experiments on the entrainment of a magnetic field by a rotating plasma, which were proposed in [24], requires the creation of a magnetic configuration of the "cusp" type. Modeling of the laser plasma transfer processes between the regions with zero field and the regions in which electrons and ions are completely magnetized involves the creation of magnetic field gradients of a level of several tesla per millimeter near the target. A zero-point cusp magnetic field configuration that meets this requirement can be created by the same pair of coils when they are connected in opposite directions.

### 3. Design of the magnetic system and test results

The main elements of the magnetic system are shown in Fig. 3. The design is based on the principles detailed in [25]. Namely, i) the solenoid was wound with a rectangular copper bar followed by making it monolithic with epoxy compound, ii) liquid nitrogen was used for cooling, iii) the energy storage recharging (recovery) circuit was rejected in order to ensure the minimum voltage between the grounded housing and the adjacent turns of the coils. Particular attention was paid to the configuration of holes in the housing that provide a plasma channel with a passage diameter of 21.5 mm in the center of the magnetic system, which opens near the edges at an angle of 11 deg to the channel axis. In the perpendicular plane, four conical channels, 12 mm wide and having aperture angles of 78 deg, were made for observation and diagnostics. The magnetic system mounting flange is located on the semispherical cover of the vacuum chamber. An insignificant displacement of the flange, and accordingly the axis of the magnetic system, with respect to the fixed axis of the laser beam propagation during evacuation can be counterbalanced by the bellows assembly, which provides positioning of the magnetic system axis within 20 mm.

      The magnetic system is composed of two Helmholtz coils wound on a stainless steel housing and is designed to operate in a single pulse mode. The pulse repetition rate is determined by the Joule loss in the coils and the time required for cooling. To reduce the ohmic resistance of the wires (about 7 times compared to room temperature) and additional thermal stabilization of the initial state, the coils are immersed in a bath with liquid nitrogen. For the magnetic system, we developed a power supply that provides the formation of a given magnetic field in the operating volume of the solenoid with inductance $L_s$ and active resistance $R_s$ when a capacitive energy storage (CES) with capacitance $C$ is discharged to it via a thyristor switch $VS$ [26]. A photograph of the power supply, its block diagram, and oscillograms of the current pulse are shown in Fig. 4. The main parameters of the power supply are summarized in Table 1 and the parameters of the coils of the magnetic system, in Table 2.

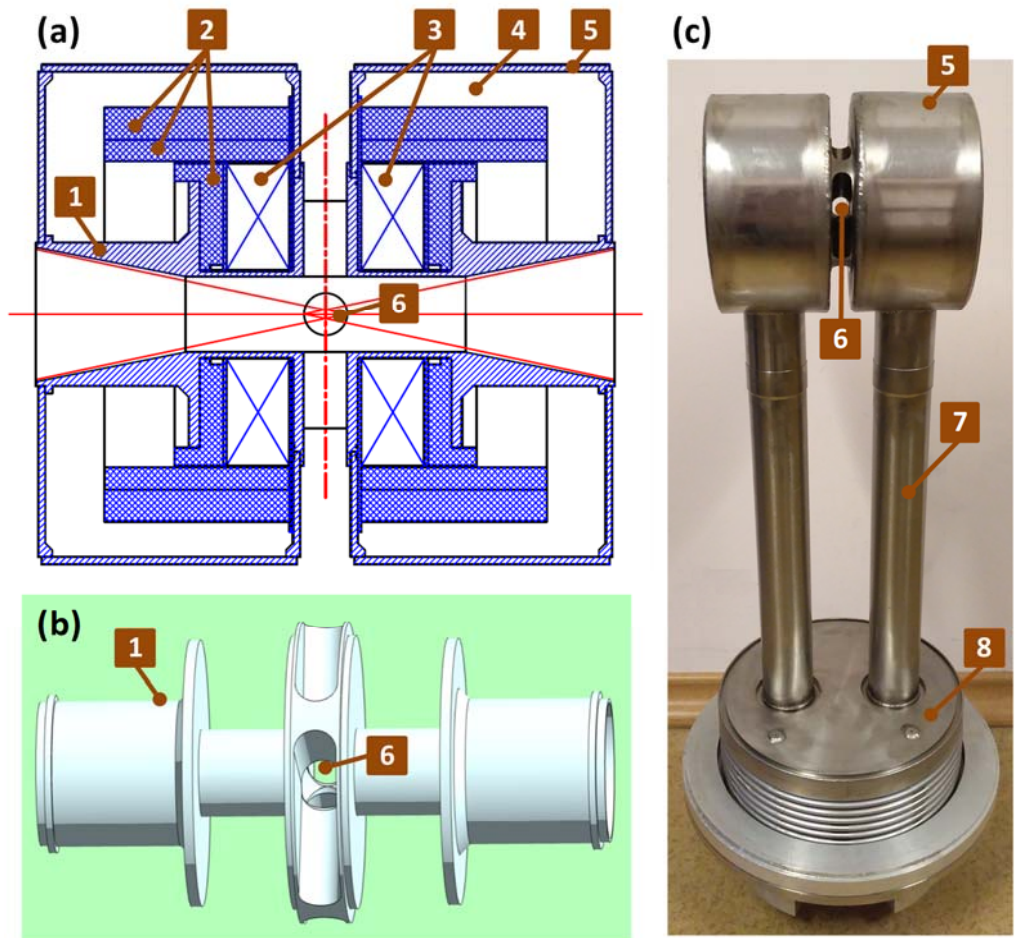

Рис.3

Fig. 3. A schematic sectional view of the magnetic system (a), a 3D model of the metal load-bearing frame (b) and the appearance of the manufactured magnetic system (c). The numbers indicate the following: 1 - metal load-bearing frame, 2 - composite elements of the load-bearing frame, 3 - windings, 4 - nitrogen chamber, 5 - external shield, 6 - conical holes for laser radiation input and plasma flow output, 7 - channels for liquid nitrogen supply and placement of current leads, and 8 - vacuum-tight bellows docking unit.

Table 1. Main parameters of the power supply of the magnetic system.

| | |
|---|---|
| Charge voltage control range | 0.2 - 3.9 kV |
| Charge voltage control discreteness | 50 V |
| Maximum admissible amplitude of the current pulse | 8 kA |
| Current pulse shape function | |
| at the leading edge and at the top | sin |
| at the rear edge | exp |
| Charge voltage instability | |
| in the range 0.2 -1.5 kV | < 1% |
| in the range 1.5 - 3.9 kV | < 0.2% |
| Charger efficiency | 90 % |

| Overall dimensions (width × depth × height) | 553×780×1270 mm |

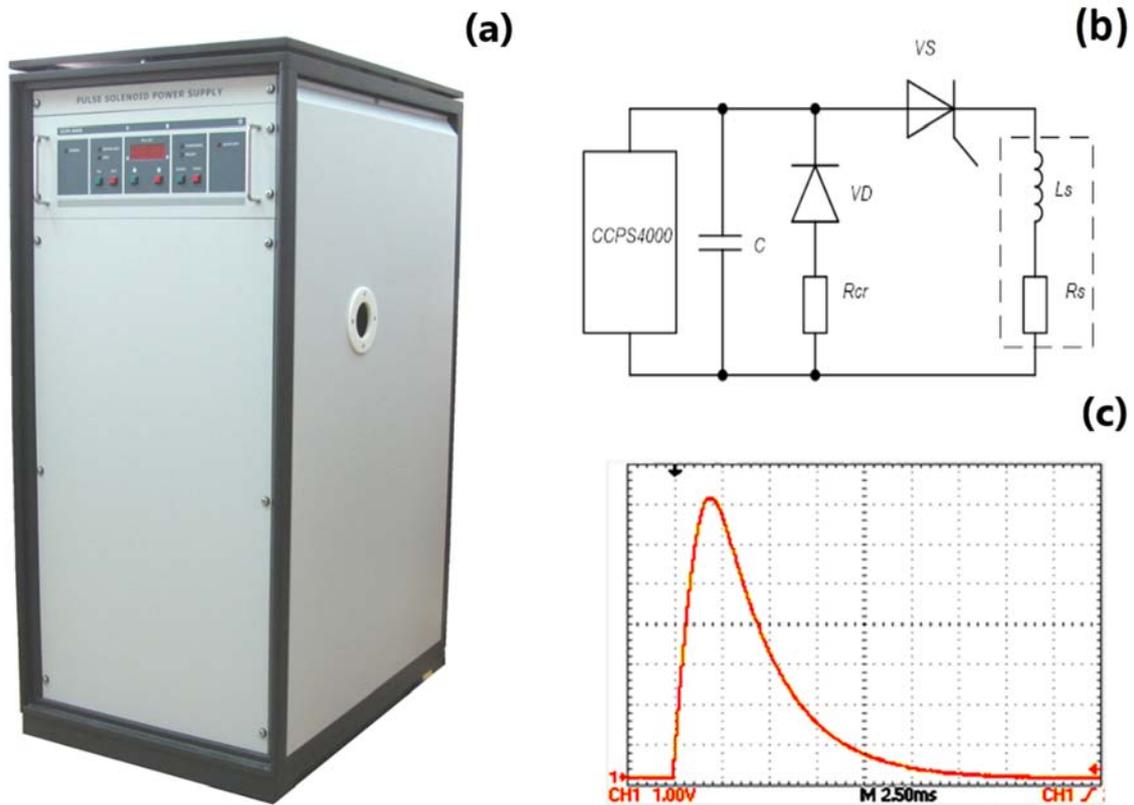

Fig. 4. Appearance of the power supply of the magnetic system (a), block diagram (b), and current pulse oscillogram (c).

Table 2. Design parameters of the magnetic system coils.

| | |
|---|---|
| Inductance of a single coil | 1.24 mH (±1%) |
| Inductance of the system with the coils connected in series | 3.67 mH (±1%) |
| Inductance of the system with the coils connected in parallel | 0.51 mH (±1%) |

The design of the magnetic system makes it possible (by switching external connections) to use oppositely connected coils or employ only one coil to significantly extend the range of possible experiments. In particular, this will permit one to study the interaction of plasma flows with non-uniform magnetic fields and the processes of separation of plasma flows from the diverging magnetic field lines.

Figure 5 shows the results of experimental measurements of the longitudinal profile of the magnetic field with in-series and oppositely connected coils, which were performed with a direct current of small amplitude using a Hall sensor. All the results meet the requirements for laser experiments, in particular, the field uniformity length in the longitudinal and radial directions is twice the characteristic size of the studied plasma (~10 mm) and completely coincides with theoretical calculations. The maximum field amplitude with co-directional connection of the coils was 15 T.

Obviously, the system experiences the greatest mechanical loads when the coils are connected in opposition. To prevent mechanical destruction, it was decided to abandon the welded connection of the parts of the winding unit, and now the latter is made by mechanical (turning and milling) processing from a single workpiece. The mechanical loads and deformations were calculated using the COMSOL Multiphysics® software package [28]. The calculation results for a 5 kA current with oppositely connected coils are shown in Fig. 6. The maximum displacement of the coil elements does not exceed 0.1 mm, and the maximum von Mises stress is 100 MPa, while the ultimate strength of the AISI 321 steel used for the manufacture of the coil frame is about 200 MPa at room temperature and only increases with cooling [29, 30].

### 4. Conclusions

A magnetic system has been created and successfully tested, which makes it possible to repeatedly generate pulses of a magnetic field with an intensity of up to 15 T. A key feature of the system is the possibility of both co-directional and oppositely directed connection of a pair of coils, which ensures a variety of magnetic field configurations.

At the moment, the operating time of the magnetic system is more than 1000 pulses, without any change in physical and technical parameters.

Using the described magnetic system, a number of successful experiments, described in, e.g., [27], were performed with the PEARL facility. In particular, laboratory modeling of astrophysical processes was carried out, and the processes of interaction of high-speed flows of hot dense laser plasma with external magnetic fields were explored to model the magnetohydrodynamic processes developing in the vicinity of compact stars. Physical processes in the boundary layer between the moving plasma and the magnetic field were examined. This is a key factor for creating physical models of the inner edge of accretion disks, accretion columns, astrophysical jets, etc.

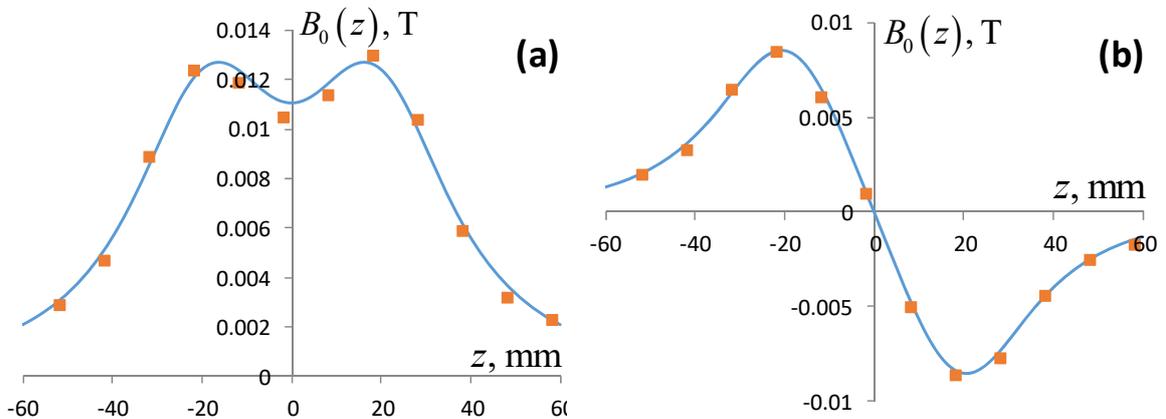

Fig.5. Distribution of the magnetic field along the system axis in the modeling mode with a 5 A current: (a) co-directional connection of coils and (b) oppositely directed connection of coils (solid lines and dots show the calculation results and the measured values, respectively).

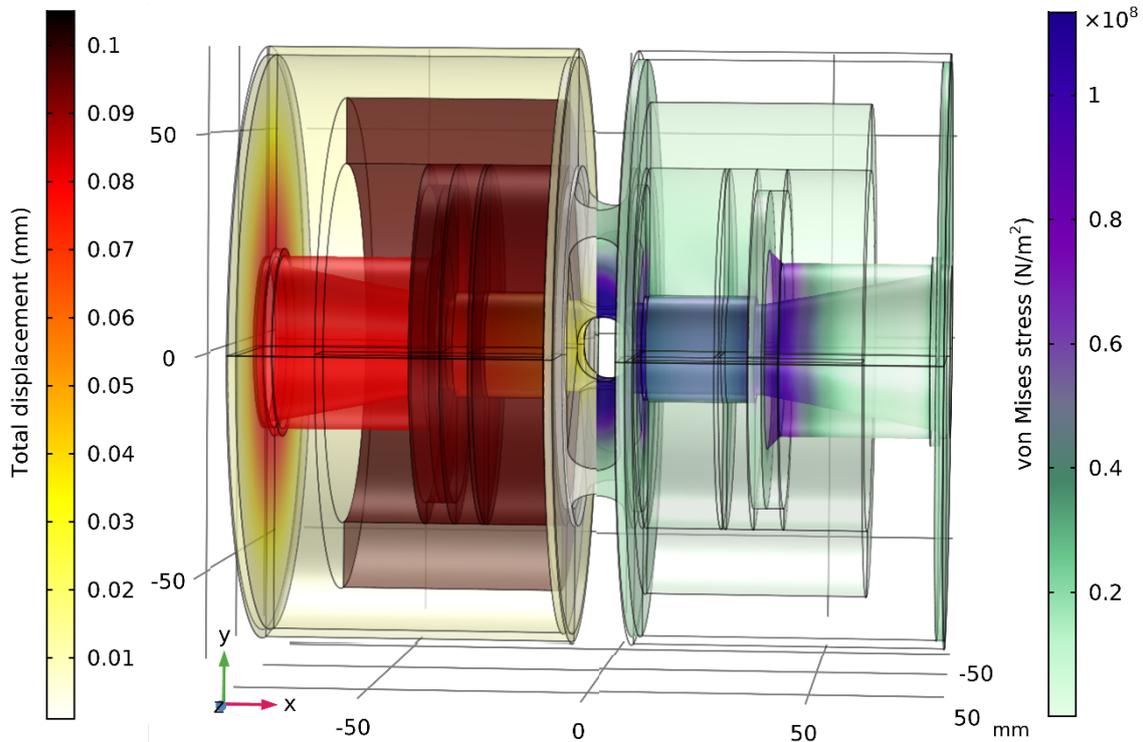

Fig. 6. Results of modeling of magnetic loads. The maximum displacement of the coil elements is shown on the left of the symmetry plane. The structure of the voltage unit for a differently directed connection of solenoids with a 5 kA current is shown on the right.

**Acknowledgements**

This work was supported by the Russian State Assignment Program, IAP RAS project 0035-2019-0001 and by the Russian Foundation for Basic Research (project no. 19-32-90102).

**Data Availability**

The data that support the findings of this study are available from the corresponding author upon reasonable request.